 \documentclass[final,5p,times,twocolumn,numeric-comp]{elsarticle}
\let\today\relax
\makeatletter
\def\ps@pprintTitle{%
    \let\@oddhead\@empty
    \let\@evenhead\@empty
    \def\@oddfoot{\footnotesize\itshape
         {Resubmitted preprint} \hfill\today}%
    \let\@evenfoot\@oddfoot
    }
\makeatother

\usepackage{amssymb}
\usepackage{siunitx}

\usepackage{listings}
\usepackage[hidelinks]{hyperref}
\usepackage{enumitem}

\newcommand{\smallsim}{$\small\sim\:\!$}

\newcommand{\refeq}[1]{Eq.~\ref{#1}}
\newcommand{\reffig}[1]{Fig.~\ref{#1}}

\journal{Computer Physics Communications}

\begin{document}

\begin{frontmatter}



\title{Advancing Scanning Probe Microscopy Simulations: A Decade of Development in Probe-Particle Models}


\author[aalto,nanolayers]{Niko Oinonen\corref{contrib}}
\affiliation[aalto]{organization={Aalto University, Department of Applied Physics},
            addressline={P.O. Box 11000 (Otakaari 1B)}, 
            city={AALTO},
            postcode={FI-00076},
            country={Finland}}
\affiliation[nanolayers]{organization={Nanolayers Research Computing Ltd},
            addressline={51 New Way Road}, 
            city={London},
            postcode={NW9 6PL},
            country={United Kingdom}}
\author[empa,marvel]{Aliaksandr V. Yakutovich\corref{contrib}}
\affiliation[empa]{
organization={nanotech@surfaces laboratory, Swiss Federal Laboratories for Materials Science and Technology (Empa)}, 
            addressline={Überlandstrasse 129}, 
            city={Dübendorf},
            postcode={CH-8600}, 
            country={Switzerland}}
\affiliation[marvel]{
    organization={National Centre for Computational Design and Discovery of Novel Materials (MARVEL), École Polytechnique Fédérale de Lausanne},
    city={Lausanne},
    postcode={CH-1015},
    country={Switzerland}
}

\author[imdea]{Aurelio Gallardo}
\affiliation[imdea]{organization={IMDEA Nanoscience Institute}, 
            addressline={C/ Faraday 9, Campus de Cantoblanco}, 
            city={Madrid},
            postcode={28049}, 
            state={},
            country={Spain}}
\author[FZU]{Martin Ondr\'a\v{c}ek}
\author[FZU]{Prokop Hapala\corref{corres}}
\affiliation[FZU]{organization={FZU - Institute of Physics of the Czech Academy of Sciences}, 
            addressline={Na Slovance 1999/2}, 
            city={Prague 8},
            postcode={182 00}, 
            state={},
            country={Czech Republic}}
\author[aalto]{Ond\v{r}ej Krej\v{c}\'i\corref{corres}}

\cortext[contrib]{Authors contributed equally}
\cortext[corres]{Corresponding authors\\ \textit{Email adresses:} \href{mailto:hapala@fzu.cz}{hapala@fzu.cz}, \href{mailto:ondrej.krejci@aalto.fi}{ondrej.krejci@aalto.fi} }

\begin{abstract}
The Probe-Particle Model combine theories designed for the simulation of scanning probe microscopy experiments, employing non-reactive, flexible tip apices to achieve \textit{sub-molecular resolution}.
In the article we present the latest version of the Probe-Particle Model implemented in the open-source \texttt{ppafm} package, highlighting substantial advancements in accuracy, computational performance, and user-friendliness. 
To demonstrate this we provide a comprehensive review of approaches for simulating non-contact Atomic Force Microscopy. They vary in complexity from simple Lennard-Jones potential to the latest {\it full density-based model}.
We compared those approaches with \textit{ab initio} calculated references, showcasing their respective merits.
All parts of the \texttt{ppafm} package have undergone acceleration by 1-2 orders of magnitude using OpenMP and OpenCL technologies.
The updated package includes an interactive graphical user interface and seamless integration into the Python ecosystem via \texttt{pip}, facilitating advanced scripting and interoperability with other software.
This adaptability positions \texttt{ppafm} as an ideal tool for high-throughput applications, including the training of machine learning models for the automatic recovery of atomic structures from nc-AFM measurements.
We envision significant potential for this application in future single-molecule analysis, synthesis, and advancements in surface science in general.
Additionally, we discuss simulations of other \textit{sub-molecular} scanning-probe imaging techniques, such as bond-resolved scanning tunnelling microscopy and kelvin probe force microscopy, all built on the robust foundation of the Probe-Particle Model.
Altogether this demonstrates the broad impact of the model across diverse domains of on-surface science and molecular chemistry.

\end{abstract}

\end{frontmatter}

\section{Introduction}

\label{introduction}

The first Scanning Tunneling Microscopy (STM) and Atomic Force Microscopy (AFM) instruments, developed in 1981~\cite{Binnig1982} and 1986~\cite{Binnig1986}, respectively, showcased the ability to visualize individual atoms of inorganic substrates.
It took, however, another two decades of scanning probe microscopy (SPM) development to distinguish individual atoms inside organic molecules separated by a distance less than~1.5~\AA {}, achieving {\it sub-molecular resolution}.
This was accomplished by passivating the apex of the metallic tip with an inert molecule (carbon monoxide, hydrogen) or atom (Xe)~\cite{Gross2009,Temirov2008}.
Due to their low reactivity, these tip apices reduce the possibility of damaging or manipulating the sample.
Furthermore, the molecules are rather loosely attached to metallic tips, which makes them flexible.
As a result, high-resolution scanning probe microscopy (HR-SPM) functions at low temperatures ($\leq 10~K$) to minimize thermal motion and prevent molecule desorption from the tip.

The flexibility of the molecule attached to the tip allows it to deflect during the interaction with the sample.
The tip apex deflection produces image distortions, which manifest themselves as either sharp lines at the ridges of the potential energy surface resembling bonds in HR-AFM~\cite{Hapala2014_PPM} or a discontinuous contrast in the HR-STM images \cite{Hapala2014_PPM,Krejci2017}.
A similar effect can be also found in Inelastic Electron Tunnelling Spectroscopy (IETS)~\cite{Hapala2014_IETS}.

SPM has also become a powerful tool for the chemical analysis and synthesis of individual organic molecules due to its ability to distinguish atoms at close distances, manipulate them, as well as to differentiate bond types.
For instance, HR-AFM with a CO-decorated tip is sensitive to the bond order in aromatic systems \cite{Gross2012}, free electron pairs in highly electronegative atoms \cite{VanderLit2016}, and the orbital configuration of transition metals in organometallic compounds \cite{DelaTorre2018}.
 
The capabilities of SPM techniques made them invaluable tools not only in fundamental research (e.g.\ for the development of futuristic molecular nanotechnology~\cite{Leinen2020} and new materials) but also in practical industrial applications. Currently, SPM helps in deciphering the chemical structures of individual molecules within complex mixtures, such as crude oil or decomposing and carbonized organic materials in the depth of oceans~\cite{Schuler2015,Fatayer2018,Kaiser2022}. HR-SPM has been also extremely useful for the recognition of complex materials and their surfaces such as calcium carbonate and fluoride \cite{Heggemann2023,Liebig2020}, showing the HR-SPM general applicability over several scientific disciplines.

Due to single molecule resolution, HR-SPM techniques allow avoiding the preparation of pure substances in macroscopic quantities which is required by other techniques for structural analysis such as X-ray or neutron diffraction.
For example, the modern AFM machines, which can employ an automatic tip preparation \cite{ALLDRITT2022108258}, are restricted mainly by the sample preparation and are physically capable of scanning thousands of molecules per day.
However, the data interpretation, typically done by teams of human experts with the aid of atomistic simulations, proves to be a tedious and challenging process.
This bottleneck hampers the broader adoption of SPM-based analytical methods beyond basic research.

The Probe-Particle Model, first introduced nearly a decade ago \cite{Hapala2014_PPM}, has become a widely used tool for simulating high-resolution AFM and STM images.
Unlike other AFM simulation models used in the fields of contact-AFM, soft-matter and biology \cite{xia2022advanced,amyot2023bioafmviewer,lopez-alonso2023pyfmlab} which focuses typically on mesoscopic aspects and AFM operation in the ambient condition, the Probe-Particle Model has been developed to explain atom-resolving non-contact AFM and STM experiments carried out in ultra-high vacuum at cryogenic temperatures with decorated tips.
The model enables the rationalization of experimentally observed SPM contrast and its attribution to chemical structure. 
The AFM part of the model, compiled into the \texttt{ppafm} computational package, is the main focus of this work.

In this specific domain, \texttt{ppafm} provides a good accuracy of simulated images at a low computational cost.
This enables rapid exploration of candidate molecular or surface structures and the exploration of suitable imaging parameters to match experimentally observed contrast to an {\it a priori} unknown geometry.
Moreover, in recent years, \texttt{ppafm} has emerged as a key driver of progress in the field of automatic interpretation of AFM data using machine-learned models \cite{Alldritt2020, CarracedoCosme2023, Tang2022}, as well as for the construction of large datasets \cite{CarracedoCosme2022} of simulated AFM data.
To the best of our knowledge, \texttt{ppafm} has served as the primary tool for generating training data for all machine-learned high-resolution AFM interpretation models published to date.

However, despite nearly a decade of development, the documentation of \texttt{ppafm} has been relatively scarce, leaving potential users largely unaware of all its features and its recent development.
Therefore, in this article, we aim to present the full spectrum of capabilities offered by the latest release of \texttt{ppafm} and present it as a comprehensive toolbox for high-throughput simulations, encompassing not only high-resolution microscopy AFM but also STM, KPFM, IETS, and other related SPM techniques:
Section~\ref{s_ncafm} describes the theoretical background of the Probe Particle Model \cite{Hapala2014_PPM} including systematic comparison of all implemented levels of theory for tip-sample interaction in order of increasing accuracy, which is missing in previous publications. There we also describe a newly implemented \textit{full density-based model} (FDBM) \cite{Ellner2016_CO} providing substantially increased accuracy.
Section~\ref{s_others} discusses models for the simulation of other SPM techniques such as STM, IETS and recently added KPFM, which builds on top of the AFM model.
Section~\ref{s_use} describes the code from the user's perspective, with the emphasis on recently simplified installation through Python Package Index (PyPI) and real-time Graphical User Interface (GUI), allowing for a friendly introduction of new users into using \texttt{ppafm}.
The technical details concerning the implementation of the method are provided in section~\ref{s_implementation}, showing the acceleration gained by smart numerical implementation, and recent parallelization on both CPU and GPU allowing for speed-up by several orders of magnitude.
Last but not least, \texttt{ppafm} is now accompanied by enhanced documentation.
We believe that these enhancements will open the field of AFM simulation towards new users and new applications in molecular design, materials science, and surface science.

\section{AFM simulation models}\label{s_ncafm}

\subsection{Tip description}\label{ss_tip}

\begin{figure}[t!]
	\centering 
	\includegraphics[width=0.5\textwidth]{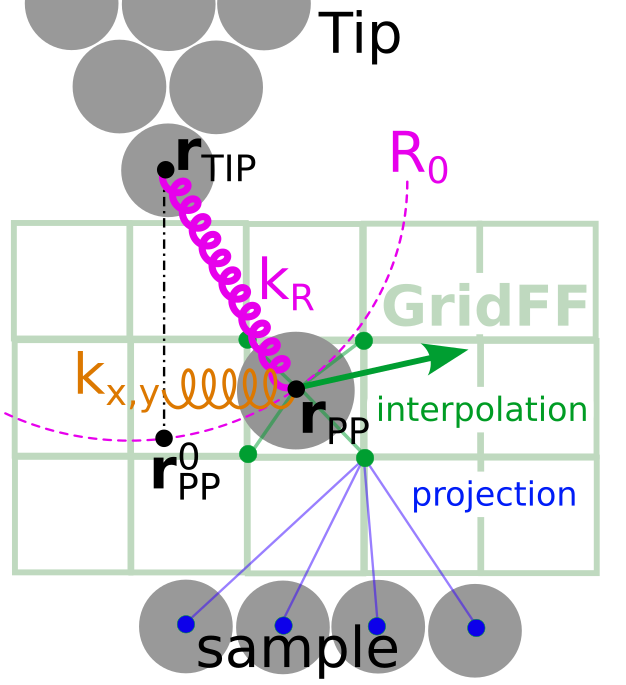}	
	\caption{\textbf{Schematics of forces acting on the flexible \textit{probe particle} (PP) in \texttt{ppafm}:}.
    The PP represents the very last atom of the non-reactive, flexible tip-apex (e.g. O atom of attached CO molecule), its position is denoted ${\vec r}_{PP}$. It is anchored to a rigid AFM tip by radial spring with high stiffness $k_R$ which keeps it in a certain distance $R_0$ from the anchor point $R_{TIP}$, and a lateral spring $k_{x,y}$ which tries to return to equilibrium position $R^0_{PP}$ under the tip. Besides the forces from the tip, also forces of the sample act on PP. These forces are calculated by interpolation of grid projected forcefield (GridFF). GridFF can be calculated by projection of atomic potentials (see \refeq{eq_LJ}, \refeq{eq_El_point}) or by convolution of grid projected densities (see \refeq{eq_El_conv}, \refeq{eq_Pauli_FDBM}).
    }
	\label{fig_tip}%
\end{figure}

The original Probe-Particle Model was based on a simple idea: simulating a non-reactive, flexible tip apex such as an attached CO molecule (or tip decoration like H$_2$, Xe \cite{Temirov2008}, NTCDA \cite{Sweetman2014} etc.) by modelling it as a single spherical particle attached to the end of an AFM tip with a spring.
This spherical particle, which we call the \textit{probe particle} (PP), represents the very last atom of the flexible tip apex (e.g. O atom in the CO-decorated tip).
This simplistic approach is motivated by the fact that the short-range forces, that determine the measured \textit{sub-molecular contrast} rapidly decay with distance and thus can be neglected for the other atoms of the tip-apex.
This allows us to separate the forces from the tip (indexed with $T$) and forces from a sample (indexed with $S$) so that the overall force  acting on the PP ($F^{PP}$) during its relaxation is evaluated as follows:
\begin{equation}
    F^{PP} ( {\vec r}_{PP}) = F^T({\vec r}_{PP}) + F^S({\vec r}_{PP}),
    \label{eq_F_PP}
\end{equation}
The forces from the sample are discussed in greater detail in section~\ref{ss_sample_tip}.

The model for the forces from the tip is as follows:
\begin{equation}
    F^T({\vec r}_{PP}) = - k_R ( |{\vec d}| - R_0) ({\vec d}/|{\vec d}|) - {\vec k_{x,y}} \odot ( {\vec d} - {\vec d}_0 ) 
    \label{eq_F_tip}
\end{equation}
Here, ${\vec d} = {\vec r}_{PP} - {\vec r}_{TIP}$ is the displacement of the PP position ${\vec r}_{PP}$ with respect to the anchor point ${\vec r}_{TIP}$ to which the PP is attached (e.g. metallic tip apex to which the CO molecule it attached).
$R_0 = |\vec{d}_0|$ stands for the equilibrium distance from the anchor point and ${\vec d}_0 = {\vec r}^0_{PP} - {\vec r}_{TIP}$ is the equilibrium displacement of PP from the anchor point (which is typically set to $(0,0,R_0)$ for a symmetric tip but may be deflected in x,y to simulate an asymmetric CO tip).
Finally, $k_R$ is the radial stiffness constant and $\vec k_{x,y}=(k_x,k_y,0)$ sets the bending stiffness and $\odot$ denotes the component-wise product of vectors.
This differs from the original model \cite{Hapala2014_PPM}, which used the Lennard-Jones potential for the radial force, keeping the PP under the tip, while here we are using the strong spring force $k_R$, as this is a computationally faster and more stable solution.
The lateral movement of the PP is controlled by the lateral springs $k_x$ and $k_y$ as it is shown in \reffig{fig_tip}(a).
From our experience, CO tips are best reproduced using a lateral stiffness of 0.24-0.25 N/m \cite{weymouth_2014}. 

The z-component of the short-range forces acting on the tip ($F_z^{tip}$) can then be calculated as the z-component of the radial spring force acting on the PP, $F_z^{tip} (\vec{r_{tip}}) = - F^T_{R,z}$ at the fully relaxed position of the PP, as these forces balance each other out.
$F_z^{tip}$ is then used for calculating the actually measured frequency shift $\Delta f$ using the formula derived by Giessibl \cite{Giessibl2001}:
\begin{equation}
    \Delta f^{tip} (\vec{r_{tip}}) = -\frac{f_0}{2k} \frac{8}{\pi A^2} \int_{-A/2}^{A/2} \frac{zF_z^{tip}(\vec{r_{tip}}+\hat{A} z)}{ \sqrt{A^2/4 - z^2} } dz,
    \label{eq_df}
\end{equation}
where $k$ is the stiffness and $f_0$ is the base oscillation frequency of the cantilever, $A$ is the peak-to-peak amplitude of the oscillation of the AFM tip and $\hat{A}$ is the normalized direction vector of the oscillation (typically in the z-direction).

\subsection{Sample-tip interaction}\label{ss_sample_tip}

The sample-tip interaction comprises of Pauli repulsion $F_{Pauli}$, van der Waals attraction (or London dispersion force) $F_{vdW}$, and electrostatic interaction $F_{el}$ between the PP and the sample:
\begin{equation}
    F^S( {\vec r}_{PP}) = F_{Pauli}({\vec r}_{PP}) + F_{vdW}({\vec r}_{PP}) + F_{el}({\vec r}_{PP}).
    \label{eq_F_sample}
\end{equation}
The precision of the \texttt{ppafm} simulation can be tuned by the level of theory describing these interactions.

In the following section, we cover the historical development of the different approximate models and their applicability.
Despite the actual implementation relying on forces, for simplicity, we only discuss formulas to compute energy components.
The respective formula for the force can be obtained as a derivative of the energy ${\vec F}({\vec r}_{PP}) =- \nabla E({\vec r}_{PP})$.

\subsubsection{Lennard-Jones}\label{ss_LJ}

In the original (and the simplest) Probe-Particle Model~\cite{Hapala2014_PPM} the motion of the PP, ${\vec r}_{PP}$, is governed by a potential obtained as a sum of pair-wise Lennard-Jones (LJ)
potentials between the PP and all the atoms of the sample.
The attractive and repulsive parts of the LJ potential simulate the attractive London dispersion and the Pauli repulsion respectively.
The total potential is evaluated as follows:

\begin{equation}
    E_{LJ}(\vec r_{PP}) = \sum_i e_{i, PP} \left[ \left( \frac{ R_{i,PP} }{ | {\vec r}_i - {\vec r}_{PP} | } \right) ^{12} - 2 \left(\frac{ R_{i,PP} }{ | {\vec r}_i - {\vec r}_{PP} | } \right)^6  \right]
    \label{eq_LJ}
\end{equation}
Here the position of the sample atoms ${\vec r}_i$ are considered rigid (i.e. not movable), and traditional mixing rules such as $R_{i, PP}=R_i+R_{PP}$ and $e_{i, PP}= \sqrt{e_i e_{PP}}$ are used to evaluate the equilibrium distance $R_{i, PP}$ and binding energy $e_{i, PP}$ of the $i$-th atom of the sample and the PP.
The default parameters $e_i, R_i$ are taken from the OPLS force field \cite{Jorgensen1988}, but \texttt{ppafm} also allows for a change of the element-based parameters in a user-provided parameter file.

\subsubsection{Lennard-Jones with point charge electrostatics}\label{ss_charges}

Shortly after the original paper~\cite{Hapala2014_PPM}, the Probe-Particle Model was modified to include the electrostatic interactions between the tip and the sample \cite{Hapala2014_IETS}.

Initially, the electrostatics was implemented as a sum of Coulomb potentials between classical point charges positioned at the centre of the PP ($q_{PP}$) and the sample atoms ($q_i$): 

\begin{equation}
    E_{el}({\vec r}_{PP}) = k_\mathrm{e} q_{PP} \sum_i \frac{ q_i }{ | {\vec r}_i - {\vec r}_{PP} | },
    \label{eq_El_point}
\end{equation}
where $k_\mathrm{e}$ is the Coulomb constant.
Simultaneously with improvements in the physics captured by the model, the assumption of a rigid sample allowed significant acceleration of the simulations.
Both the electrostatic and the LJ force field are pre-calculated and stored on a real space grid, from which they are interpolated during the simulations, as illustrated in Fig.~\ref{fig_tip} and explained in sec.~\ref{tech_implement}.

\subsubsection{Lennard-Jones with density functional theory based electrostatics}\label{ss_dft_electro}

A more accurate model of the electrostatics was developed further, using a grid-based real-space representation of the electrostatic potential of the sample ($ V_{S}$).
$V_{S}$ is obtained as the Hartree potential from sample electronic structure calculation in density functional theory (DFT). 
The electrostatic potential acting on the PP with its charge density ($\rho_{PP}$) is obtained through a cross-correlation integral:
\begin{equation}
    E_{el}({\vec r}_{PP}) = \int_{\vec r} \rho_{PP}({\vec r}
    ) V_{S}(\vec r +{\vec r}_{PP}) d\vec{r}.
    \label{eq_El_conv}
\end{equation}

We found that the distortions in AFM images by electrostatic field to a large extent explain for example the over-enhanced bond-length contrast in fullerenes or other Kekule structures \cite{Gross2012,Hapala2015_book} but also the repulsive contrast over triple bonds and free electron pairs \cite{VanderLit2016}.
Nevertheless, the charge required to reproduce experimental contrast with monopole charge distribution was unrealistically high (0.2-0.4e). 

In further applications \cite{Hapala2016,Peng2018NatComm,Peng2018Nature} we concluded that the quadrupolar charge distribution better reproduces contrast observed with a CO-tip.
The quadrupolar charge distribution is better fitting the CO molecule and the CO-tip charge density as concluded by DFT calculations \cite{Peng2018NatComm,Ellner2016_CO}.

While point-charge electrostatics proved useful for quick and easy model calculations independent of ab initio inputs, which were often conducted by external experimental groups through a web interface \cite{ppafm_web}, DFT-based electrostatics of the sample was, nevertheless, found necessary to properly simulate intricate image effects, such as those arising from free electron pairs and triple bonds.
For the CO tip the quadrupole moment can vary in between -0.025 to -0.15 $e\times\AA{}^2$, depending on the experiment \cite{DelaTorre2017,Alldritt2020,Hapala2016}.

A minor disadvantage of the cross-correlation-based approach (see sec.~\ref{tech_implement}) is the assumption that the PP moves without rotation.
However, according to our experience with the complex-tip model \cite{Giovannantonio2018} and comparison of the Probe-Particle Model against the direct integration model by Ellner et al.~\cite{Ellner2016_CO} the differences caused by the multipole rotation are minor.
This can be understood from the fact that bending angles are rather small at tip-sample distance relevant for high-resolution imaging experiments, and the bending is most significant at the close range where the interaction is dominated by the Pauli rather than electrostatic interaction.

\subsubsection{Full density-based model}\label{ss_FDBM}

Pauli repulsion modeled by the repulsive part $(1/r^{12})$ of the spherically symmetric LJ potential cannot reproduce delicate effects emerging from rearrangements of the electron density in the sample which are often visible using HR-AFM techniques \cite{Gross2012,VanderLit2016,DelaTorre2018}. Some of these limitations can be mitigated by modification of the LJ parameters of individual atoms (especially van der Waals radius) to match the iso-surface of electron density obtained from a DFT calculation \cite{DelaTorre2018}. This approach allows to distinguish between different occupations of the atomic orbitals for atoms of the same element and it was also successfully used for calculations of ionic materials, such as calcite or calcium fluoride \cite{Liebig2020,Heggemann2023}. Nevertheless, such approach is still limited by the spherical symmetry of the LJ potential, therefore it cannot fully recover non-spherical effects such as free-electron pairs and variation of density in covalent bonds.

In order to addresses these limitations, Ellner et al.\ \cite{Ellner2019} introduced an improved model called the \emph{full density-based model} (FDBM), where both the Pauli repulsion and electrostatics are calculated directly from electron density obtained from DFT. While electrostatics is still calculated using \refeq{eq_El_conv}, the Pauli repulsion is newly calculated by the integral of the product of the tip and the sample charge densities scaled by a fitting constant $A$.
Eventually the product is raised to exponent $\beta$ (although $\beta$ is typically close to one): 
\begin{equation}
    E_{Pauli}({\vec r}_{PP}) = A \int_{\vec r} \left[  \rho_{PP}({\vec r} 
    ) \rho_{S}({\vec r}+{\vec r}_{PP} ) ) \right]^{\beta} d\vec{r}
    \label{eq_Pauli_FDBM}
\end{equation}
The magnitude of the repulsion is significantly more sensitive to the exponent $\beta$ than the multiplicative factor $A$. Even a change of only $0.1$ in $\beta$ results in a significant change in the observed contrast, higher values typically resulting in reduced sharpness. However, if the scanning distance and $A$ are adjusted along with $\beta$, similar-looking contrast can be observed with multiple distinct combinations of the parameters.

The resulting model, combined with an appropriate dispersion interaction model (previously modeled by the attractive part of the LJ potential), and properly fitted, could remarkably reproduce experimentally measured images of rigid molecules. Particularly, it better captures the free electron pairs (e.g., oxygen and nitrogen heteroatoms) and Kekule structures (e.g., triple bonds), which were previously only emulated through the repulsive electrostatic field in the original LJ-based model sometimes using unrealistically high tip charge \cite{Hapala2015_book}. Now FDBM also accounts for Pauli repulsion, capturing the electron hardness of free electron pairs on oxygen and nitrogen atoms.

The dispersion interaction model typically used with the FDBM is the Grimme DFT-D3 \cite{DFTD3params} dispersion correction, which we have also recently implemented in \texttt{ppafm}, in particular in the Becke-Johnson damping form \cite{BJ-damping}.
One notable aspect of the DFT-D3 correction is that the interaction coefficients for each atom depend on their chemical environment, based on proximity, to account for the changing polarizability due to bonding. In principle, the distance calculations to determine the bonding configuration would also include the PP. However, since the PP is supposed to be chemically inert, we choose to exclude the PP from this calculation, which allows the interaction coefficients in the sample to be calculated independently of the PP position, significantly speeding up the calculation.

The DFT-D3 energy also has parameters that are adjusted for particular DFT functionals - namely \textit{s$_6$}, \textit{s$_8$}, \textit{a$_1$}, and \textit{a$_2$} \cite{DFTD3params}.
So far there has not been any extensive study on the effect of these parameters on \texttt{ppafm} simulations and thus we recommend sticking to the parameters connected with the DFT functional used for the calculation of the FDBM model input. 
\texttt{ppafm} provides here predefined parameter values for many commonly used DFT functionals.

\subsubsection{Comparison of AFM simulation models}\label{ss_afm_discussion}

\begin{figure*}[h!]
\centering
\includegraphics[width=0.9\textwidth]{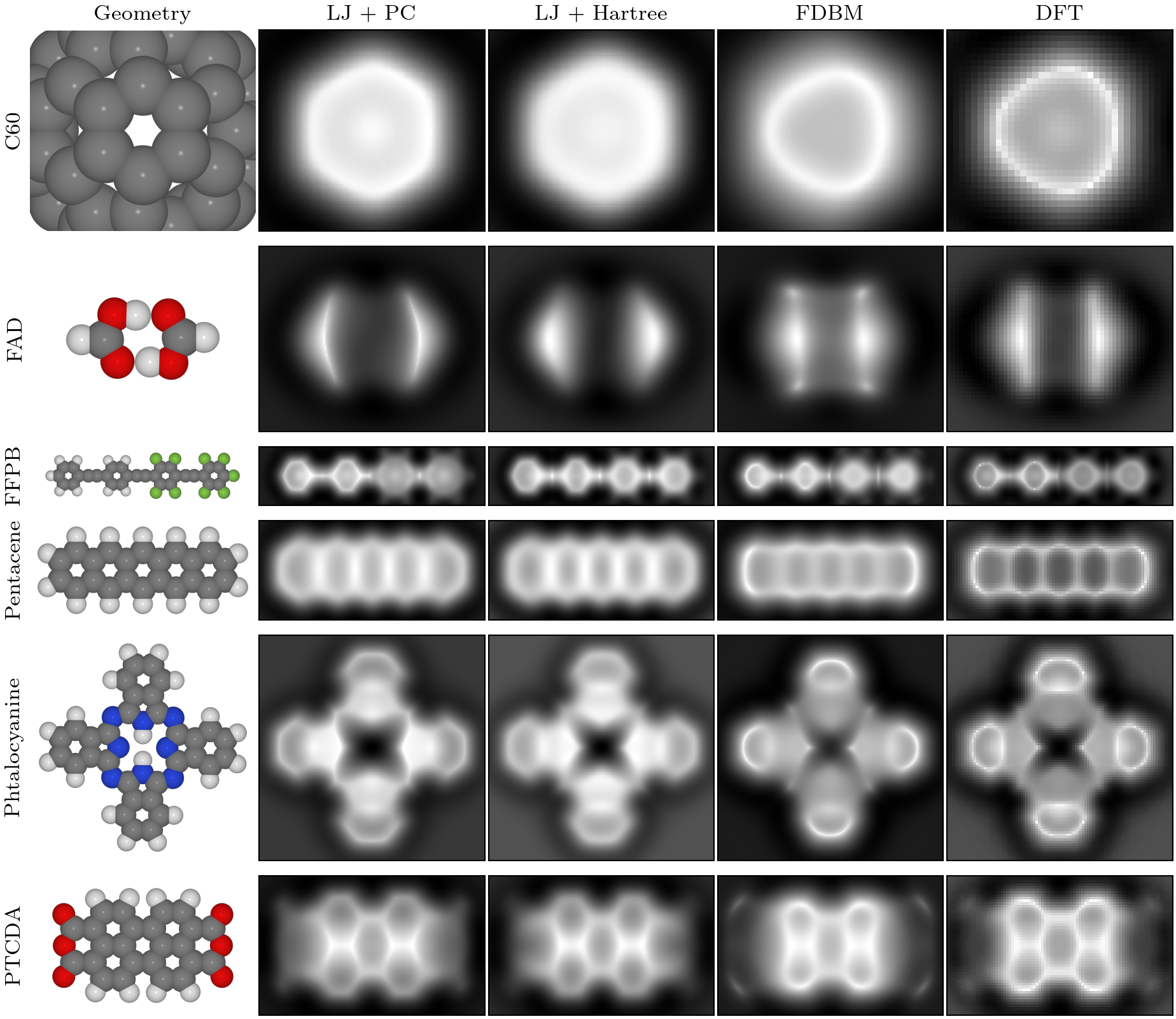}
\caption{\textbf{Comparison of \texttt{ppafm} simulations with different models of tip-sample interaction.} The columns depict different models. Starting from the left, they are Lennard-Jones + point charges (LJ+PC), Lennard-Jones + Hartree (LJ+Hartree), Full density based model (FDBM), and a DFT simulation used as a reference, where the tip-sample interaction at each pixel was calculated as an independent relaxation of the tip using the CP2K program.
For the Lennard-Jones based simulation model (both with point-charge and Hartree potential) we used a quadrupole charge distributions on the tip with the quadrupole moment $-0.05\, e\times\AA{}^2$. For the point-charge simulations we used Mulliken charges reported by FHI-aims \cite{Blum2009}. 
For FDBM model we set the multiplicative factor $A = 12$ and exponent $\beta = 1.2$ (see \refeq{eq_Pauli_FDBM}) which provided the best overall match to DFT for all the molecules. } 
\label{fig_AFMmodels}
\end{figure*}

To illustrate the strengths and weaknesses of each tip-sample interaction model we plot in \reffig{fig_AFMmodels} results calculated for a representative selection of molecules.
We compare \texttt{ppafm} simulations against DFT reference calculations performed using CP2K \cite{cp2k} with PBE exchange-correlation functional and Grimme DFT-D3 \cite{DFTD3params} used for the van der Waals correction.
Each of the selected molecules represents some characteristic chemical moieties manifested as characteristic features in AFM and was previously discussed in HR-AFM-related literature. To make the comparison consistent we choose the same simulation parameters for each molecule, even though this choice may not be optimal to represent the DFT reference or experimentally observed contrast.
This means that for the FDBM model, we set the multiplicative factor $A = 12$ and exponent $\beta = 1.2$ (see \refeq{eq_Pauli_FDBM}), which provided the best overall match to DFT for all the molecules.
We found that the best match is visible if we offset z-distance by $-0.2\, \AA$ between the DFT and the \texttt{ppafm} simulations to get a roughly matching level of sharpness in the observed contrast.
In the FDBM simulation, the electrostatics used a DFT-calculated charge density on the CO-tip. For the Lennard-Jones-based simulation model (both with point-charge and Hartree potential) we used quadrupole charge distributions on the tip with quadrupole moment $-0.05\, e\times\AA{}^2$.
For point charges simulations we used Mulliken charges reported by FHI-aims \cite{Blum2009}.

{\bf C60 Fullerene} was studied as an example of bond order discrimination \cite{Gross2012}. The difference in the apparent bond length, as well as the ovaloid shape of the electron cloud, is very well reproduced by the FDBM model.
To some degree the difference in apparent bond length can be reproduced also with the LJ+Hartree model, nevertheless unrealistically high charge of the tip is needed to reproduce experimental (or DFT calculated) contrast \cite{Hapala2015_book}.  

{\bf FAD} (Formic acid dimer) represents carboxylic groups which often dimerize in self-assembled structures studied by AFM \cite{VanderLit2016,Zahl2021}. The FDBM again provides contrast most similar to DFT data, including bright spots above oxygen atoms. This is due to the ability of FDBM to reflect localized electron pairs in Pauli repulsion. Nevertheless, these bright spots are visible also in LJ+PC and LJ+Hartree simulations at higher tip-sample separation where electrostatic forces dominate \cite{VanderLit2016}. 

{\bf FFPB} molecule (4-(4-(2,3,4,5,6- pentafluorophenylethynyl)- 2,3,5,6- tetrafluorophenylethynyl) phenylethynylbenzene) was studied to see the effect of electron depletion on the AFM contrast in a benzene ring ($\pi$-hole), caused by the electron-withdrawing substituents (fluorines) \cite{Moll2014,Mallada2023}.
In DFT simulations this is visible as darker contrast over fluorinated rings, which can be attributed to a faster decay of the electron density \cite{Moll2012} (due to deeper electrostatic potential and lower work function) and by electrostatic attraction between the ($\pi$-hole) and free electron pair of the CO tip.
Surprisingly, this effect is best reproduced by the LJ+PC model, which used Mulliken charges obtained from DFT calculation.
Another characteristic feature is the triple bond rendered as a bright line perpendicular to the bond.
This effect is caused by the toroidal shape of the $\pi$-electron cloud around the triple bond \cite{deOteyza2013,VanderLit2016}, which produces a quadrupolar field both in electrostatics and Pauli repulsion. FDBM again reproduce this feature best thanks to the incorporation of proper aspherical Pauli repulsion, while the Lennard-Jones potential is composed of spherical potentials around each atom and therefore cannot reproduce this feature.
Nevertheless, both LJ+PC and LJ+Hartree can reproduce the electrostatic contribution of this repulsive feature.

{\bf Pentacene} molecule was one of the first molecules for which bond-resolved AFM images were measured \cite{Gross2009}. Besides the five hexagonal rings, the experiment and DFT simulations show increased repulsion over the ends of the aromatic system. This effect is to a large degree caused by a higher attractive van der Waals background in the center as was explained in the original paper \cite{Gross2009}. Nevertheless, our simulation done at closer tip-sample separation shows, that the effect is pronounced even at a distance where van der Waals contribution is negligible. This is reproduced by FDBM but not with LJ-based models. Without detailed analysis, we can only speculate that this is because tails of occupied frontier molecular orbitals (HOMO, HOMO-1 etc.) which contribute most to Pauli repulsion are more suppressed in the center due to the presence of nodes.
All models including FDBM and LJ-based models reproduce very well the elongation of the rings perpendicular to the molecule axis, which is caused by deflection of the probe mostly due to the lateral gradient of van der Waals potential (with a slight contribution from electrostatics), as was discussed previously \cite{Neu2014,Hapala2016}.

{\bf Phtalocyanine} molecule was widely studied in the SPM community \cite{Liljeroth2007,Mohn2012,DelaTorre2018,ChenPhtalo2023} due to its great potential for molecular electronics, catalysis, and for biological importance of porphyrin derivatives. The main features which can be seen in HR-AFM experiments and which are perfectly reproduced by DFT simulations are (i) The bright peripheral benzene rings contrasting against the darker porphyrin center, and (ii) sharp pointy corners of imine nitrogens. Both of these features are nicely reproduced by FDBM, which properly accounts for the Pauli repulsion affected by slower decay of electron clouds in benzene rings (with respect to the porphyrin center), as well as Pauli repulsion of the free electron pairs of these nitrogens. The LJ-based model incorrectly renders the pentagonal rings brighter.
This is a simple effect of a higher concentration of repulsive atoms in the pentagon ring in contrast to the hexagon, as the LJ model cannot account for the rate of decay of tails of electron density. Nevertheless, the pointiness of the nitrogen groups is rather well reproduced mostly due to the significant role of the electrostatic forces which cause the apparent shrinking of these areas as previously discussed \cite{Hapala2014_IETS}.   

{\bf PTCDA} (Perylenetetracarboxylic dianhydride) is perhaps the most studied molecule in the SPM community \cite{Temirov2008,Moll2012,Hapala2014_PPM,Hapala2016,Dolezal2021}, mostly due to experimental convenience and formation of well ordered self-assembled monolayers. The experiments as well as DFT simulations show the central perylene system considerably brighter than the peripheral anhydride groups. This is more-or-less reproduced by all models, although the FDBM model excels in this aspect, as it reflects higher Pauli repulsion due to the longer extent of the electron cloud over the perylene system \cite{Moll2012}. All models properly describe the apparent enlargement of the anhydride groups and shrinking of the perylene group caused by electrostatic forces \cite{Hapala2016}. In addition, the DFT simulation shows bright repulsive features over the carbonyl oxygens, which are again best reproduced by the FDBM model.      

Despite the generally superior accuracy of the FDBM approach, the \texttt{ppafm} code allows users to choose from various simulation models (Lennard-Jones, Morse, point charges, model charge density integration, FDBM) the one which offers an optimal compromise between accuracy and simplicity for their particular application. Such a choice should not be motivated by the computational cost of AFM simulations, as our efficient GPU implementation allows interactive simulations even on the FDBM level.

Nevertheless, the simpler models (e.g. LJ + point charges) limit reliance on DFT data (i.e. charge density and Hartree potential are not required).
This makes those simple models very convenient for fast screening over various modeled sample geometries or the creation of databases for machine learning approaches.
The dependence of the DFT-based electrostatics and FDBM method on DFT calculations (at least a thousand times slower) and a large amount of volumetric data are making this method less attractive for fast high-throughput simulation scenarios.
For rapid training of AFM recognition models, we recommend pre-training the model on the data obtained from simple LJ and point-charge-based simulations, with the refinement step performed on fewer examples generated by the FDBM (similar approach was used in \cite{Tang2022}).

Notice that LJ simulations presented in \ref{fig_AFMmodels} were done using default LJ parameters which depend only on chemical elements, not on more detailed atomic types (i.e.\ we do not distinguish different sub-types of carbon like sp$^1$, sp$^2$, aromatic, carboxylic etc.). With a more careful selection of atomic types and of LJ parameters (particularly the atomic radius) even simple LJ model can simulate the different extents of electron clouds and bring the LJ-PC model closer to AFM experiments without the need for DFT inputs \cite{DelaTorre2018,Liebig2020}. Although the FDBM model does not depend on such a detailed choice of atomic types (assuming van der Waals D3 parameters are given, and have a minor effect on resulting contrast), it still depends on the choice of the two global parameters (scaling factors $A$ and exponent $\beta$ in \refeq{eq_Pauli_FDBM}). The optimal choice of these two parameters is still under debate and may be system dependent.

\section{Other PP-SPM simulation modes}\label{s_others}

The interaction models discussed in the previous section are the central part of the Probe-Particle Model as they determine forces acting on the PP and therefore also its deflection. This deflection then modifies the measured contrast of other signals (such as STM \cite{Hapala2014_PPM, Krejci2017} and IETS \cite{Hapala2014_IETS}), typically by sharpening or introducing discontinuities to the contrast.
In addition, other forces can emerge in the junction between tip and sample e.g. due to polarization of the PP or the molecule under study by an external electric field.
These microscopic contributions of the polarization force responsible for the \textit{sub-molecular} contrast in Kelvin probe force microscopy (KPFM) can also be simulated with the \texttt{ppafm} package.
KPFM and other simulation techniques built on top of the original model are discussed in this section.


\subsection{Kelvin probe force microscopy} \label{ss_KPFM}
\begin{figure}[t!]
    \centering
\includegraphics[width=0.9\linewidth]{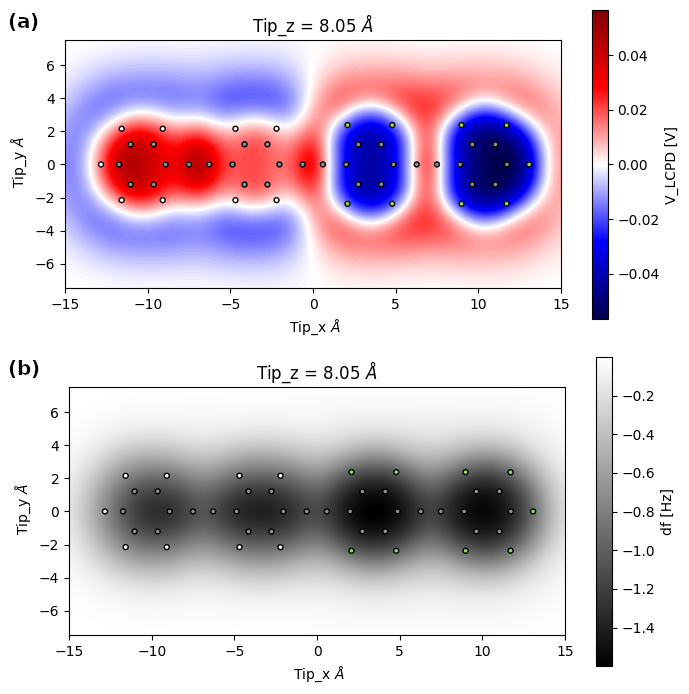}

    \caption{(a) LCPD map taken over the FFPB molecule, simulated using the KPFM functionality of \texttt{ppafm}. (b) AFM image for the same tip distance as in the LCPD map.
    Images in both panels are overlaid with the atomic structure of the FFPB molecule.
    The Tip\_z = $8.05$~\AA{} distance quoted as the height of the scanning plane position was measured between the molecular plane and the anchor pivot (metallic apex) of the tip. The tip distance $z_\mathrm{tip} = 4.05$~\AA{} used to scale the electric field induced by the voltage between the tip and the molecule was smaller by $R_0 = 4.00$~\AA{}. A CO-tip model with the static quadrupole moment of $-0.05$~$e\times\AA{}^2$ and default electric polarizability was used. The capacitance of the metallic part of the tip was modeled with a sphere of the $R_\mathrm{tip} = 40$~nm radius.}
    \label{fig_KPFM}
\end{figure}

Traditionally, KPFM experiments measured the electrostatic forces between tip and sample due to external electric potentials and differences between the work functions of the two materials. In this process, the tip and the sample can be seen as the plates of a capacitor.
The force between such plates depends quadratically on the potential difference between the tip and the sample $V$ and linearly on the gradient of the effective capacitance $C({\vec r}_{tip})$ with respect to the position of the tip.
\begin{equation}
    {\vec F}({\vec r}_{tip})=\frac{V^2}{2} \nabla C({\vec r}_{tip}).
    \label{eq_F_KPFM}
\end{equation}

Although KPFM experiments were originally intended to measure mesoscopic features such as the work function of the studied materials and long-range charge domains, the development of atomically precise SPM techniques had allowed to achieve \textit{sub-molecular} KPFM contrast, corresponding to variations of the charge distribution and polarizability within individual organic molecules \cite{Mohn2012,Albrecht2015,Moll2014,Schuler2014}.
Nevertheless, the quantitative relation between the measured quantities and the electronic structure of the molecules was under debate.
We introduced a KPFM module into the \texttt{ppafm} code to put these relations on quantitative ground and provide a straightforward tool for the simulation of these phenomena.

In this implementation, the bias dependence of both the charge density of the probe and the electrostatic potential of the sample is introduced in \refeq{eq_El_conv}, to study its effect on the force ${\vec F}_{PP}({\vec r}_{tip})$ and the corresponding frequency shift $\Delta f$. As has been shown in our previous publications \cite{Mallada2021,Mallada2023}, the \textit{sub-molecular} variation of $\Delta f(V)$ originates mostly from the intrinsic charges within the tip or the sample that interact with bias-induced electric polarization of the opposite electrode.
The output of the KPFM mode can be represented as a map of (apparent) \emph{local contact potential difference} (LCPD or $V_\mathrm{LCPD}$), which corresponds to the bias voltage at which the maximum of the (approximately) parabolic $\Delta f(V)$ dependence lies.

Currently, the KPFM functionality is implemented in the \texttt{ppafm} package in two variants. In the first version, the changes in the charge densities of the tip and sample due to the application of an external field in the z-direction must be provided as inputs from external DFT calculations. In the second version, analytically generated tip polarizations, fitted to the DFT calculated ones, are provided for user convenience. For a more detailed description of the usage of the KPFM module and the theoretical basis of the model, please refer to the code manual and the supplementary information of \cite{Mallada2021}.

As an example of a KPFM simulation, \reffig{fig_KPFM}a shows the LCPD map over the FFPB molecule. The LCPD is affected by the local electric charges on the molecule: positive charge under the tip tends to shift the LCPD towards more negative values, and negative charge towards more positive values. The resulting figure clearly shows the polarization within the molecule. The two positive-charged (electron-depleted) benzene rings in the right-hand-side half of the molecule are surrounded by negative charge of the fluorine atoms, while the two electron-rich benzene rings in the left-hand-side half are surrounded by more positive hydrogen atoms. Notice that the simulated contrast is consistent with experimentally measured KPFM pictures from literature \cite{Moll2014}. KPFM experiments usually need to be performed with larger tip--sample distance as compared to HR-AFM, resulting in a much more blurred AFM image (\reffig{fig_KPFM}b) in comparison with the HR-AFM of FFPB in Fig.~\ref{fig_AFMmodels}.

\subsection{Bond-resolved STM}\label{ss_tmp}

Despite the fact that the bond-resolved STM technique preceded \textit{sub-molecular resolution} in {AFM \cite{Temirov2008}, this technique received less attention in scientific community because the interpretation of the measured signal was unclear. In order to put the interpretation of these techniques on more quantitative grounds, and provide a straightforward simulation tool, we developed \texttt{ppstm} \cite{Krejci2017,ppstm_repository} which builds on top of \texttt{ppafm}. The \texttt{ppstm} code can be used as a standalone STM simulation package (independent of \texttt{ppafm}) to simulate normal STM with rigid (e.g.\ metallic) tip. It is based on Chen's rules \cite{Chen_1990} approximation of Bardeen tunneling theory to evaluate tunneling current between the tip and sample.

Nevertheless, the real strength of the \texttt{ppstm} code is in its capability to calculate high-resolution (i.e.\ bond-resolved) STM images obtained with flexible tip-apices (e.g.\ CO, Xe, H$_2$).
In this application \texttt{ppstm} is combined with the \texttt{ppafm} code for pre-calculating the PP positions ${\vec r}_{PP}$ for each position of the tip ${\vec r}_{tip}$ during the scanning.
This essentially just shifts the position of the orbitals located on the PP involved in the Chen's tunneling formulas, as is illustrated in~\reffig{fig_ppstm}(a), which distorts the resulting image and gives rise to the characteristic sharp contrast in the high-resolution STM images as was described in \cite{Hapala2014_PPM,Krejci2017}.
Bond-resolved imaging and STM simulations can be especially beneficial for STM machines without AFM possibilities, since STM measurements are tipically experimentally simpler than AFM. This was used for example for study of carbon nanoribbons and other graphitic structures \cite{Rizzo2018,Lawrence2020}.

The main drawback of this approach is the added complexity of the interpretation and theoretical rationalization of the measured STM signal, which depends both on geometrical relaxation of the PP as well as on precise estimation of elusive electronic structure of the sample, and the tip~\cite{Krejci2017}, as is illustrated  in~\reffig{fig_ppstm}(b) and (c).
Precise hybrid functional DFT calculations of the whole sample (i.e. including both molecule and substrate) are often necessary in order to achieve agreement with experiment \cite{DelaTorre2017}.

\begin{figure}[t!]
	\centering 
	\includegraphics[width=0.5\textwidth]{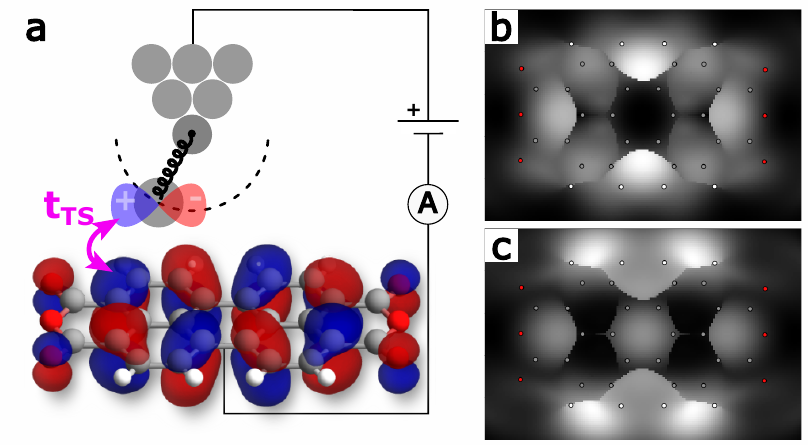}	
	\caption{\textbf{Schematics of \texttt{ppstm} with example of PTCDA}. 
   (a) \texttt{ppstm} calculates the electron tunneling rate $T_{TS}$ between the molecular orbitals of the sample and the tip using Chen's rules \cite{Chen_1990}. The orbitals of the tip are modeled by atomic orbitals, like $s, p_z, p_x$ (on the image) and $p_y$, positioned at the probe particle. This is done for the PP positions previously relaxed by \texttt{ppafm} in order to account for the displacement of e.g. a CO molecule caused by the interaction with the sample. (b,c) Examples of a simulated \texttt{ppstm} of the PTCDA molecule image using the \textit{dIdV} mode. The model electronic structure of the tip was chosen to be $13\%$ of $s$ PP orbital and $87\%$ of $p_x$ and $p_y$ orbitals. This configuration best reproduced experimentally observed contrast measured with CO-tip in the previous publications (e.g.~\cite{DelaTorre2017}). (b) at the sample bias corresponding to the HOMO orbital energy. (c) at the sample bias corresponding to the LUMO orbital energy. Please note that in the experiment the STM contrast can be affected by interaction of molecular orbitals with the interface states of the substrate, forming a delicate electronic structure.} 
\label{fig_ppstm}%
\end{figure}

\subsection{Inelastic scanning tunneling microscopy}\label{ss_iets}

\begin{figure}[t!]
\centering 
\includegraphics[width=0.5\textwidth]{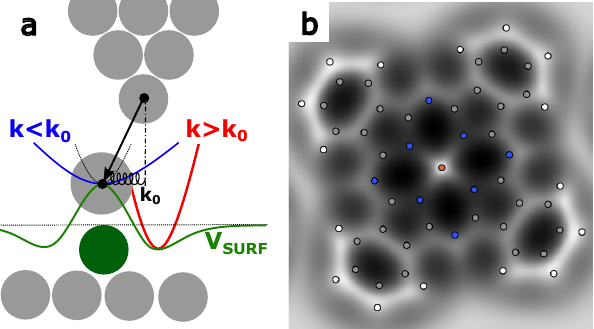}	
\caption{\textbf{IETS simulation encoded in the \texttt{ppafm} with example of Iron Phtalocyanine (FePc)}. 
{\bf (a)} Schematic illustration of the IETS imaging mechanism as built in the \texttt{ppafm}. The stiffness of the lateral CO vibration ($k_0$) is modified by curvature of the tip-sample interaction potential $V_{surf}$. The stiffness as well as the associated vibration frequency is increased at the position of a convex $V_{surf}$ and decreased at position of a concave potential (e.g. above atoms and bonds). Due to broadening of peaks measured in IETS spectroscopy, this softening of the modes above atoms and bonds allows detecting higher amplitude of inelastic tunneling signal originating from lateral CO vibration below the base energy $\epsilon_0$ associated with the base stiffness $k_0$, as was done by the Ho group \cite{Chiang2014}. For a more detailed explanation please see \cite{Hapala2014_IETS,DelaTorre2017}. {\bf (b)} Example of an IETS map calculated by \texttt{ppafm} on the iron phtalocyanine (FePc) molecule using just a Lennard-Jones force-field. Notice that in this particular simulation neither the distortions caused by electrostatics as discussed in \cite{Hapala2014_IETS} nor the modulation of the tunneling signal by orbital symmetry discussed in \cite{DelaTorre2017} are present.}
\label{fig_IETS}
\end{figure}

Another method to achieve \textit{sub-molecular resolution}, very close to the HR-AFM contrast, was demonstrated by the Ho group with inelastic STM \cite{Chiang2014}; however, without explanation of its mechanism. Already in the same year we were able to explain and simulate the observed contrast with the new IETS module added to our \texttt{ppafm} package \cite{Hapala2014_IETS}. This module calculates the change of the stiffness (resp.\ vibration frequency) of lateral vibration modes of the CO molecule attached to the AFM tip due to its interaction with the sample (see \reffig{fig_IETS}a). In repulsive regime the ridge-lines in tip-sample interaction potential (e.g.\ over the bonds between atoms in the sample) introduce negative curvature to the total potential in which the CO molecule vibrates. This effectively decreases the stiffness and vibration frequency of the relevant vibration mode, therefore shifting the inelastic tunneling peaks to lower energy.
This effect is visible in simulated map shown in \reffig{fig_IETS}b, where bright contrast above the atoms and bonds correspond to increased inelastic tunnelling signal at energy (i.e. bias voltage set-point) below the base energy of lateral CO vibration mode. The amplitude of the peak is related to shift of the peak energy when considering e.g. Gaussian broadening as explained in \cite{Hapala2014_IETS}. 

Later we were able to improve this technique 
through considering the variation of the tunneling current, which depends on the orbital symmetry \cite{DelaTorre2017}.
The inelastic signal modulation via the electron-phonon coupling between the tunneling (calculated by Chen's rules as in section~\ref{ss_tmp}) and the lateral vibration mode, are described in detail in \cite{DelaTorre2017}. 

\section{\texttt{ppafm} from the user perspective}\label{s_use}

\subsection{Installation}\label{ss_install}

The default way to install the \texttt{ppafm} code is through the \texttt{pip} installer for Python packages.
This can be achieved by running
\begin{lstlisting}
    pip install ppafm
\end{lstlisting}
on the command line.
This installs the package from the Python Package Index (PyPI), which contains pre-compiled distributions of the \texttt{ppafm} code for several operating systems.
If the binary files for your environment do not exist, \texttt{pip} will attempt to compile them upon download.
Additionally, \texttt{pip} installs all the necessary Python dependencies.
Some non-Python GPU dependencies, however, might be installed separately from appropriate OpenCL-capable GPU drivers and related libraries.
For more experienced users and developers we provide alternative ways of installing and running the code.
The most up-to-date installation instructions can always be found in the repository \cite{ppafm_repository}.
They include installation in a dedicated Conda environment, Docker container, building from the source code, and others.

\subsection{Command-line user interface} \label{ss_cmd}

The command-line interface (CLI) of the \texttt{ppafm} code provides users with access to the full capabilities of the package.
This is an alternative to the graphical user interface discussed in the following section.
The CLI interface allows to run simulations on supercomputers, cloud computers, and other computational resources without a graphical interface.
Also, the CLI interface is used for high-throughput simulations when run by a workflow manager.

Once the \texttt{ppafm} package is installed user gets access to a variety of tools to compute force fields, relax the probe particle, and plot the results.
Below is an example of launching \texttt{ppafm} to compute the Lennard-Jones force field:
\begin{lstlisting}
    ppafm-generate-ljff -i structure.xyz
\end{lstlisting}
In the example above we specify an XYZ file with an input structure.
Additionally, the user can create a "params.ini" file containing more fine-tuning settings of the simulation.
If the file is present in the folder, the code will pick it up automatically.
The \texttt{ppafm} repository \cite{ppafm_repository} contains detailed instructions on how to use \texttt{ppafm} through the CLI.

\subsection{Graphical user interface} \label{ss_gui}

\begin{figure}[t!]
	\centering 
	\includegraphics[width=0.5\textwidth]{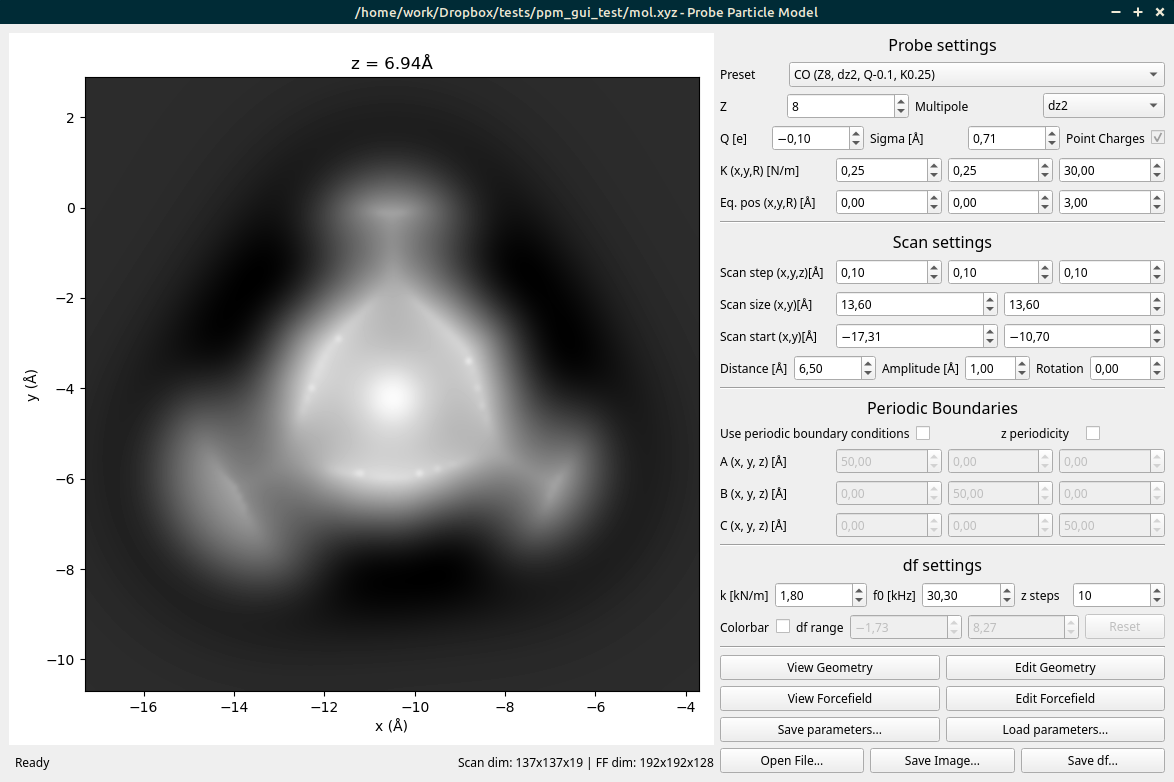}	
	\caption{Interactive GUI for GPU-accelerated \texttt{ppafm} simulations. The user can interactively modify imaging parameters by using input boxes or a mouse-wheel, and image contrast is updated automatically. } 
	\label{fig_GUI}%
\end{figure}

The latest GPU-accelerated version of the \texttt{ppafm} code is so fast that interacting with the user solely through scripts or a bash terminal becomes a significant bottleneck. Simulations of a full stack of AFM images with typical resolutions of 200x200x20 pixels take \smallsim\SI{0.1}{s} on a typical desktop computer equipped with a dedicated GPU. For this reason, we have developed a simple graphical user interface (GUI) (see \reffig{fig_GUI}) that enables users to quickly explore simulation results obtained with different inputs. Users can vary parameters such as the bending stiffness of the CO tip, oscillation amplitude of the AFM cantilever, effective charge of the tip, or parameters of the FDBM model, and immediately visualize the results for comparison with experimental references. This approach is particularly useful for new users who are trying to familiarize themselves with the code and gain intuition about how various imaging parameters can affect measured AFM contrast. This exploration can be also useful to the experimentalist who wishes to gain an idea of what to expect from the images from a real HR-AFM machine. Another application is manually finding the set of parameters that most closely resembles a specific set of reference AFM images obtained with a particular setup. This can be used, for example, in the refinement of training data for machine-learned models of automatic image interpretation.
\subsection{Integration with other software} \label{ss_integration}

In order to fully exploit the computational efficiency of the \texttt{ppafm} code (especially in GPU-accelerated version) for machine-learning and other high-throughput applications, we provide a Python application programming interface (API) which allows for seamless integration with other Python-based software. In particular, this API was used to rapidly generate training data for a machine-learning application for automated AFM image interpretation \cite{Alldritt2020}.

The Python API is structured in multiple levels that reflect the different computational steps in the \texttt{ppafm} simulation. On the high level, the user can simply provide a molecular geometry or Hartree potential from a DFT calculation, construct a simulator with given physical parameters, and run the whole simulation in one step. On the lower level, the user could choose to manually construct the PP-sample force field with the different chosen force field models (sec. \ref{ss_sample_tip}) or run the PP relaxation for a given force field.

The API also provides tools for easily creating large datasets of AFM simulations for machine learning applications. In particular, this was used in a previous study by Alldritt et al.~\cite{Alldritt2020}, who used so-called image descriptors for identifying the atomic structures of molecules.
\texttt{ppafm} provides implementations for several different image descriptors that the user can compute for a given molecular geometry. In order to create datasets of both AFM simulations and any desired image descriptors, we provide a high-level generator API that takes a list of samples (geometry, Hartree potential) as an input and generates batches of samples containing the AFM images, the descriptors, and the molecule structures, ready for use in machine-learning training as is or for storing on the disk for later use.
Additionally, it is possible to introduce randomizations to the simulation parameters during the generation process to account for parameters varying during the experiment, either from a predefined list of randomization operations or custom user-defined operations.

\section{Implementation Details} \label{s_implementation}

\subsection{Numerical methods} \label{tech_implement}

\subsubsection{Grid force-field} \label{tech_imp_grid}

In order to accelerate the relaxation the PP interacting with the sample we split the simulation into two steps: 

\begin{enumerate}[listparindent=1.25em, parsep=0pt]

\item \textbf{Force field generation}: The first step involves projection of all components of the sample potential and force-field (i.e. electrostatic, Pauli, van der Waals, see \refeq{eq_F_sample}) onto a uniform rectangular real-space grid covering the whole simulation supercell. This means, that for each such grid point we sum atomic contribution in Eqs.~\ref{eq_LJ} and \ref{eq_El_point} from all atoms of the sample, or evaluate the integrals in Eqs.~\ref{eq_El_conv} and \ref{eq_Pauli_FDBM}. 

A typical spacing of the grid points is 0.1-0.2 Å which produces 1-10 million sampling points for typical simulation supercell of size 20x20x20 \AA{}. In the CPU implementation the components of these grid force-fields are saved into files (e.g. FFLJ\_$[x|y|z]$.xsf, FFel\_$[x|y|z]$.xsf). In the GPU implementation this is ommited, since saving and loading of these data files from disk is often slower than the evaluation on the GPU. 
    
\item \textbf{Relaxation}: In the second step, the PP position is optimized by the FIRE relaxation algorithm \cite{Bitzek2006} using the forces interpolated from previously constructed grid force field. Currently we used tri-linear interpolation of the forces (which corresponds to quadratic interpolation of the potential). But we are experimenting with tri-cubic interpolation of the potential which may allow us to use larger grid spacing and avoid storage of forces (i.e.\ improve memory efficiency).
    
\end{enumerate}

In practice, the two-step simulation procedure was found to be approximately 10-100 times faster than implementation not using an intermediate grid-based force field for typical samples comprising of tens to hundreds of atoms. The simulation speed of the two-step procedure is typically limited by the first step (force field generation), which takes roughly 1 minute on a single CPU for typical grid size comprising of a million points (100x100x100). In the case of Lennard-Jones and point-charge electrostatics the algorithm is perfectly parallelizable and it scales proportionally to number of CPUs when OpenMP acceleration is used (which is on by default in the CPU version) and it takes just \smallsim\SI{0.1}{s} when using OpenCL accelerated code on contemporary GPU equipped desktops with thousands of cores.

\subsubsection{Convolution theorem} \label{tech_imp_conv}

The evaluation of the electrostatic force-field from the electrostatic potential of the sample and the tip charged density distribution \refeq{eq_El_conv} and the evaluation of the Pauli repulsion from the overlap of the sample and tip charged densities \refeq{eq_Pauli_FDBM} have the form of a cross-correlation. Therefore they can be expressed using the convolution theorem simply as a product in the Fourier space (with an additional complex conjugation in the cross-correlation case). For a typical grid size (e.g. 100x100x100 = 1 million points) such transformation using Fast Fourier transform is orders magnitude faster than direct integration of the formulas \refeq{eq_El_conv}, \refeq{eq_Pauli_FDBM} point-by-point in real space (the scaling is $O( (n \log(n))^3)$ for FFT vs $O(n^6)$ for direct integration, where $n=100$ is the grid dimension in one direction). The calculation of \refeq{eq_El_conv} and \refeq{eq_Pauli_FDBM} using FFT was implemented on both CPU and GPU and the computational cost is similar to Lennard-Jones and point-charge electrostatics. In the current implementation the CPU version computes the FFT using NumPy \cite{Harris2020} (not parallized) and the GPU version uses Reikna \cite{reikna_repository}.

\subsection{ Code structure} \label{ss_code}

\subsubsection{Python package with a C++/OpenCL backend} \label{tech_imp_python}

\texttt{ppafm} code is designed to behave as standard Python package and exposes a Python front-end to the user, allowing sophisticated scripting. Python (with NumPy) is used to implement of the high-level logic, and most of utility functions for saving and loading simulation parameters, molecular geometry and some operations on 3D datagrids. Matplotlib library is used for plotting of final results. The computational core of the package is implemented in C++ (for CPU version) and OpenCL (for GPU version). The C++ code is interfaced with Python using the ctypes-library in Python.

\subsubsection{GPU implementation} \label{tech_imp_gpu}

Modern graphics processing units (GPUs) possess thousands of independent computing cores, offering orders of magnitude higher raw computing power than traditional CPUs. However, efficient utilization of this computing power is limited to tasks that are naturally parallel (i.e., independent) and not memory-bound (either by main memory bandwidth or cache size). AFM simulations are ideal for GPU acceleration since the simulations of individual pixels (i.e., positions of the AFM tip) are virtually independent. Furthermore, the simulation scheme that evaluates the sample potential through interpolation of the real-space grid can be accelerated using texture interpolation hardware. Therefore we ported all performance intensive tasks on GPU using pyOpenCL. This includes the FFT convolution, projection of atom-wise LJ, electrostatic, and D3 van der Waals force fields to the grid, and relaxation of the probe particle position). Generally speaking, the GPU accelerated simulations are so fast that the timing is relevant only for interactive work (GUI) or high-throughput tasks such as machine learning.


\subsection{Performance} \label{ss_code}

\begin{figure}[t!]
	\centering 
	\includegraphics[width=0.5\textwidth]{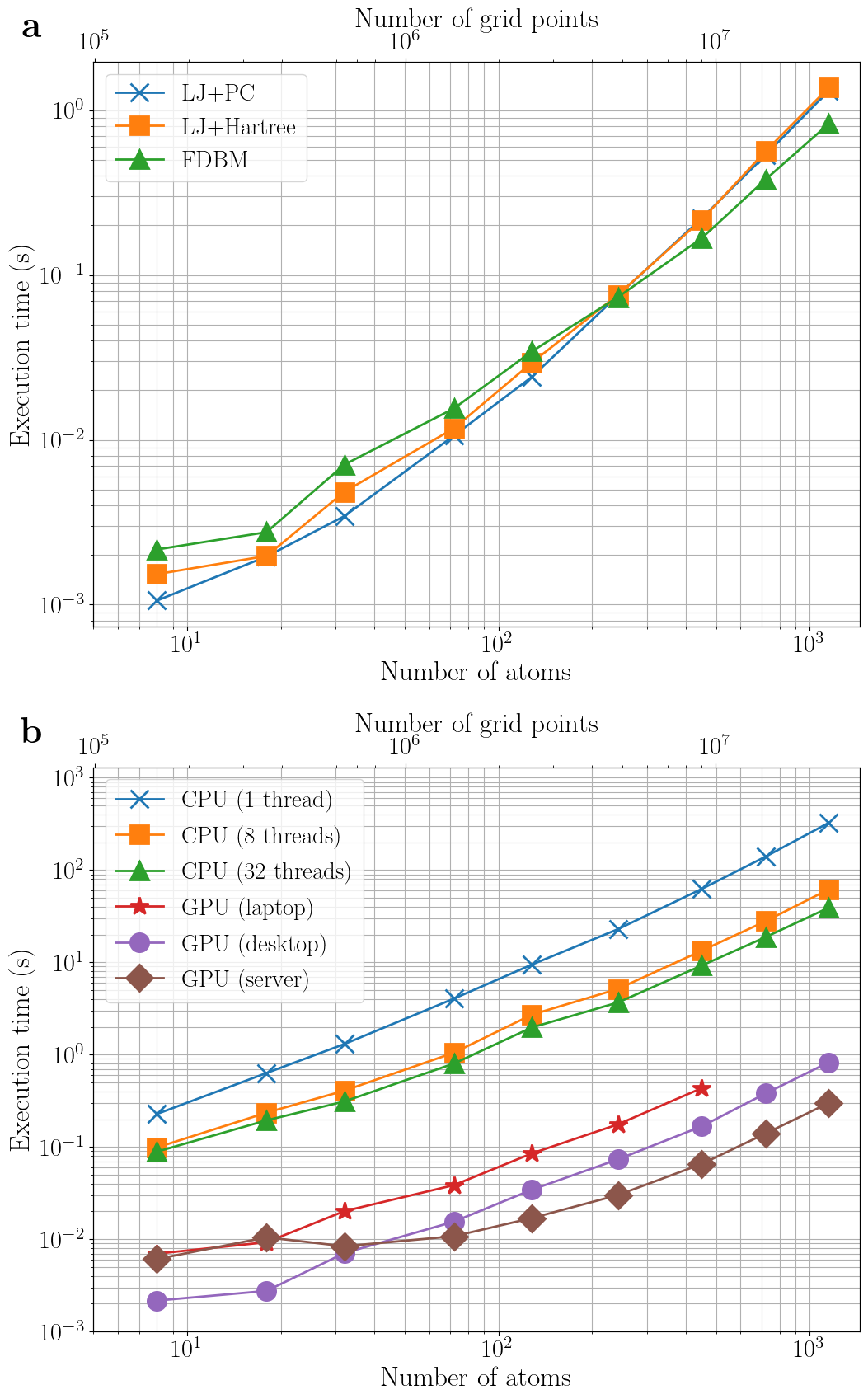}	
	\caption{
    \textbf{Examples of performance scaling of the \texttt{ppafm} code on various hardware}.
    Results shown in both plots (a, b) were measured for simulation of a box containing a periodic graphene sheet of different sizes and thus number of atoms.
    Notice that the number of computational grid points is proportional to the number of atoms, which allows us to plot the results with a common x-axis, in this example.
    (a) The total execution time (i.e. both force-field generation and relaxation) of GPU accelerated simulation, for three different levels of theory depending on the size of the system:
    Lennard-Jones potential with point-charges electrostatics (LJ + PC), Lennard-Jones potential + electrostatics from the Hartree potential (LJ+Hartree) and the full density based method (FDBM).
    These calculations were proceeded on AMD RX 6700 XT GPU.
    (b) Comparison of the force-field generation step for FDBM method using OpenMP acceleration with multi-core CPU and various GPUs with OpenCL.
    The used CPU is Intel i9-13900K, and the GPUs are:
    laptop Nvidia GTX 1650Ti Mobile, desktop AMD RX 6700 XT and server Nvidia A100.
    The last two points for the laptop GPU are missing because the memory requirement exceeds the capability of the GPU.
    }
     \label{fig_speed}%
\end{figure}

We conducted thorough performance tests (see \reffig{fig_speed}) on a periodic (infinite) graphene sheet using different levels of theory (see section \ref{ss_sample_tip}) and hardware (both CPU and GPU).
The simplicity of this test system allows us to systematically scale the simulation size in a broad range by varying the size of the simulation box, when both the number of atoms $n$ and grid points $m$ scales proportionally at the same time.
Therefore we can plot both on the same x-axis.
Figure \ref{fig_speed}a shows the comparison of total simulation time of different methods (Lennard-Jones + point charges (LJ+PC), Lennard-Jones + Hartree (LJ+Hartree), and full density-based model (FDBM)) on a desktop GPU.
Notice the comparable performance cost and rather ideal scaling for all methods over the whole range of system sizes.
The exceptions are the smallest systems ($<$10 atoms), where initialization and other overheads become the bottleneck.
For small systems the LJ+PC method is \smallsim2x faster than FDBM, while for a system with $>$200 atoms FDBM actually becomes cheaper. 

This can be rationalized by different asymptotic scaling of the algorithms.
While the projection of the atomic force field on the grid used for calculating the Lennard-Jones and point-charge electrostatics scales as $O(nm)$, the product of grid size $m$ and the number $n$ of atoms, the FFT-based cross-correlation used to compute the Hartree and Pauli potentials in FDBM scales as $O(m log(m))$ with the grid size $m$ and is notably independent of the number of atoms $n$.
The LJ+Hartree method using a combination of both algorithms is in between LJ+PC and FDBM method.
It should be noted that execution time relationship could vary for different types of systems, e.g.\ a non-periodic system with empty space on the sides of the simulation box. 
Such systems can have a significantly smaller number of atoms, while having a similar grid size.
This would lead to the LJ+PC method being significantly faster than the other two.
The relaxation of the PP, which scales as $O(m)$, typically takes a negligible share ($<$5\%) of the total simulation time, which is dominated by the force-field generation.

To clearly demonstrate the speedup achieved with GPU (OpenCL) and CPU (OpenMP) parallelization, we also compare the time required to build the FDBM force field on different platforms (see \reffig{fig_speed}b): a CPU (Intel i9-13900K) using a varying number of threads, a laptop GPU (Nvidia GTX 1650Ti Mobile), a desktop GPU (AMD RX 6700 XT), and a server GPU (Nvidia A100). The measured performance profiles demonstrate that the OpenCL implementation even on the laptop GPU outperforms a single-core CPU by roughly two orders of magnitude over the whole range of sizes.
The exceptions are the largest systems which do not fit into the laptop GPU memory. The server GPU is by yet another order of magnitude faster, except smallest sizes where the performance is limited by initialization overheads in the server environment. Notice that the CPU performance for 32 threads is not proportionally improved in comparison to 8 threads. 
This is because the CPU code uses NumPy's implementation of FFT, which is not affected by OpenMP settings.
Replacing the NumPy FFT by a different better-parallelized FFT library is a simple way to further improve performance on CPU in the future.

These measurements exclude additional time taken by loading the input files from disk and preparing arrays in the GPU memory, which actually become the bottleneck for single small simulations. However, these operations are amortized if a batch of simulations is run using the same grid, as is the case for example in the GUI when changing simulation parameters unrelated to the grid size.

\section{Summary and conclusions}\label{s_conclusion}

In this paper, we summarize the significant development that Probe-Particle Model has gone through during its roughly decade-long history since its inception \cite{Hapala2014_PPM}, and illustrate its computational efficiency together with its versatility through wide variety of applications in the field of high-resolution scanning probe microscopy.
The sub-molecular AFM simulations model, compiled into the \texttt{ppafm} package, integrates different levels of theory, described in section \ref{ss_sample_tip}, allowing to balance speed and accuracy and analyze the effect of different physical interactions on the resulting AFM image.
Even the most accurate method implemented (FDBM) cannot exactly reproduce all physical interactions between the tip and the sample (such as deformation of electron clouds and displacement of sample atoms), which can be captured by expensive quantum mechanics calculations like DFT.
Nevertheless, as demonstrated by \reffig{fig_AFMmodels}, it can typically match all relevant features extremely well at a tiny fraction of the computational cost.
This efficiency and the user-friendly interface, through the command-line or the GUI, makes \texttt{ppafm} an ideal tool for quickly searching over different sample structures often used for sample structure recovery.
The unparalleled numerical performance of \texttt{ppafm} (especially in its GPU-accelerated version) has been recently exploited for the production of large databases of simulated AFM data for training machine-learned models for the reconstruction of molecular geometries from AFM images.
In this area we expect great application potential, as it opens door to widespread use of high-resolution SPM methods as tool for routine single-molecule analysis.

\section*{Acknowledgements}

O.\ K.\ wants to thank to Adam S.\ Foster and Patrick Rinke for discussions and support.
N.\ O.\ has been supported by the World Premier International Research Center Initiative (WPI), MEXT, Japan and by the Academy of Finland (Projects No.\ 347319, 347611, 346824).
A.\ V.\ Y.\ acknowledges the NCCR MARVEL funded by the Swiss National Science Foundation (grant No.\ 205602).
A.\ G.\ acknowledges the financial support from the "Juan de la Cierva" fellowship (JDC2022-048249-I
).
P.\ H.\ gratefully acknowledges the financial support by the Czech Science Foundation Junior Star project 22-06008M.
O.\ K.\ has been supported by the European Union’s Horizon 2020 research and innovation programme under the Marie Skłodowska-Curie grant agreement No.\ 845060.
The authors gratefully acknowledge Czech computer infrastructure Metacentrum, CSC – IT Center for Science, Finland, and the Aalto Science-IT project for the generous computational resources.  Metacentrum resources are provided by the e-INFRA CZ project (ID:90254), supported by the Ministry of Education, Youth and Sports of the Czech Republic.

\section*{Data availability}

All the data and computational procedures for obtaining the simulated images presented in this paper can be found in \cite{zenodo}.

\appendix

\bibliographystyle{elsarticle-harv}
\bibliography{resubmission_arXiv/paper}






\end{document}